\documentclass[aps,prl,twocolumn,showpacs,groupedaddress]{revtex4}
\usepackage{graphicx}

\begin{document}


\title{A Molecular Matter-Wave Amplifier}
\author{Chris P. Search
and Pierre Meystre}
\affiliation{Optical Sciences Center, The
University of Arizona, Tucson, AZ 85721}

\date{\today}

\begin{abstract} We describe a matter-wave amplifier for
vibrational ground state molecules, which uses a Feshbach
resonance to first form quasi-bound molecules starting from an
atomic Bose-Einstein condensate. The quasi-bound molecules are
then driven into their stable vibrational ground state via a
two-photon Raman transition inside an optical cavity. The
transition from the quasi-bound state to the electronically
excited state is driven by a classical field. Amplification of
ground state molecules is then achieved by using a strongly damped
cavity mode for the transition from the electronically excited
molecules to the molecular ground state. \end{abstract}

\maketitle

Recent progress in the application of Feshbach resonances to
ultracold atomic gases has led to the production of macroscopic
numbers of ultracold molecular dimers starting from either a
Bose-Einstein condensate  (BEC) \cite{donley} or a degenerate
Fermi gas \cite{regal1}. This work has culminated in the
production of molecular Bose-Einstein condensates starting from a
Fermi gas of $^{40}$K \cite{greiner} and $^6$Li \cite{zwierlein}.
The ability to coherently produce a macroscopic number of quantum
degenerate molecules opens up new avenues of research in the field
of matter-wave optics such as molecular interferometry, the
molecular laser, and nonlinear wave mixing between atomic and
molecular fields. In fact, the effective Hamiltonians that
describe the coherent atom-molecule conversion are formally the
same as second harmonic generation in nonlinear optics with a
$\chi^{(2)}$ nonlinear susceptibility.

A major stumbling block on the road to coherent molecular optics
is the short lifetime of the molecules formed via Feshbach
resonances. While these molecules are translationally very cold,
they are vibrationally hot and can decay to lower lying
vibrational states via inelastic collisions with other atoms and
molecules. For an atomic BEC, the molecules are expected to decay
in a time $\sim 10-100\mu s$ for typical atomic densities
\cite{yurovsky}. For atomic Fermi gases, recent results indicate
that the molecular lifetime can be several orders of magnitude
longer due to the underlying Pauli blocking between atoms
\cite{petrov}.

Still it would be desirable to create molecules in their
vibrational ground state so as to maximize the amount of time with
which experiments can be done with them. This can in principle be
done using two-photon stimulated Raman photoassociation
\cite{wynar}. This is a relatively inefficient process due to the
very small Frank-Condon factors between the free atom pairs and
the molecules. This is a consequence of the small spatial overlap
between the tightly bound molecular wave function and the large
average separation between pairs of atoms in these dilute gases.
One way to overcome this difficulty is to use a Feshbach resonance
to form weakly bound molecules, followed by a two-photon Raman
transition to transfer the molecules to a low lying vibrational
state \cite{kokkelmans}. The advantages of this procedure are
two-fold. First, recent experiments with Feshbach resonances in
quantum degenerate gases have demonstrated the ability to readily
convert a significant fraction of the atoms into molecules.
Secondly, the spatial overlap between the resulting weakly bound
states and the ground-state molecules is substantially larger than
for that of the free atom pairs, which results in larger
two-photon transition rates. Indeed, several experiments using
single-photon photoassociation spectroscopy have observed a strong
enhancement of the photoassociation rate in the presence of a
Feshbach resonance \cite{courtielle}.

In this paper we use the idea of Feshbach stimulated
photoproduction \cite{kokkelmans} as the basis for a matter-wave
amplifier for ground-state molecules. Active elements such as
matter-wave amplifiers are an essential element for the field of
matter-wave optics and have previously been demonstrated
experimentally for atoms \cite{inouye}.

In our scheme, pairs of atoms in a BEC are coupled to a
quasi-bound state of a closed collisional channel via a Feshbach
resonance \cite{timmermans}. A two-photon transition couples the
quasi-bound molecular state to the vibrational ground state of the
molecules. A large amplitude classical electromagnetic field
drives the the quasi-bound state to an electronically excited
molecular state, followed by the emission of a photon into a
single mode of a quantized optical cavity field. This second step
takes the molecule from the excited state to its vibrational and
electronic ground state. By using a strongly damped cavity one can
achieve continuous unidirectional coherent amplification of an
initial molecular signal field that persists until the atomic
condensate is depleted \cite{law}. Although the number of photons
in the cavity field is less than one at all times, strong coupling
of the molecules to the cavity field can be achieved by
controlling the volume and geometry of the cavity as was recently
demonstrated in the optical regime with the single atom laser
\cite{mckeever}.

The discrete mode structure of the cavity allows one to select a
particular vibrational state of the electronic ground-state
molecules provided the cavity linewidth, $\kappa$, is less than
the vibrational level spacing, $\delta \nu\sim 1$GHz. This is a
distinct advantage over using free space spontaneous emission to
initially populate the ground state, as in atomic matter-wave
amplifiers \cite{inouye}, since spontaneous emission would
populate a large number of vibrational levels of the ground state
as well as leading to dissociation back into the continuum
\cite{nikolov}. As a result, a molecular matter-wave amplifier
relying on free space spontaneous emission would require that the
initial signal field that is to be amplified be very large in
order that stimulated scattering dominates over spontaneous decay
into other bound vibrational and continuum states.

Atoms in the BEC are assumed to all occupy the same hyperfine
state denoted by $|0\rangle$. Pairs of atoms in $|0\rangle$ are
coupled to the quasi-bound molecular state $|1\rangle$ via a
Feshbach resonance with coupling strength $\alpha$ and with an
energy difference between pairs of zero energy atoms and the
quasi-bound state given by the difference in the Zeeman
interaction energies for the two states,
$\omega_1$\cite{timmermans}. State $|1\rangle$ is coupled to an
electronically excited molecular state, $|2\rangle$, via a
classical optical field with Rabi frequency $\Omega_l$ and
frequency $\omega_l$. State $|2\rangle$ is also coupled via a single mode
of the cavity field with vacuum Rabi frequency $g$ and frequency $\omega_c$ to
molecules in their electronic and vibrational ground state, $|3
\rangle$.  The internal energies of states $|2\rangle$ and
$|3\rangle$ relative to pairs of atoms in $|0\rangle$ are
$\omega_2$ and $\omega_3$, respectively. We assume that
$\Delta=(\omega_2-\omega_1)-\omega_l\approx
(\omega_2-\omega_3)-\omega_c \gg |g|,|\Omega_l|,\gamma_e$ where
$\gamma_e^{-1}$ is the lifetime of $|2\rangle$. Under these
conditions the excited state can be adiabatically eliminated,
leading to two-photon Raman transitions between $|1\rangle$ and
$|3\rangle$ with the two-photon Rabi frequency $\chi(x)=g
u(x)\Omega^*_l(x)/\Delta$ where $u(x)$ is the mode function of the
cavity field.

We assume that the atoms and the weakly bound molecules in state
$|1\rangle$ are confined by a trapping potential inside the
optical cavity. The state to be amplified, $|3\rangle$, may also be confined
by a trapping potential or molecules in that state may pass through
the cavity as a wave packet\cite{law}. Here we only treat explicitly the case of molecules
in $|3\rangle$ being in a stationary state of their center-of-mass motion.
At zero temperature we can use a single mode approximation for the
atomic and molecular fields with the center-of-mass wave functions
$\phi_n(x)$ and energies $\epsilon_n$ for $|n=0,1,2,3\rangle$. The
Hamiltonian, in a rotating frame in which the energy of
$|3\rangle$ is given by the two-photon detuning,
$\delta=(\omega_3+\epsilon_3)-(\omega_1+\epsilon_1)-(\omega_l-\omega_c)$,
consists of three terms,
$H=\delta\hat{b}^{\dagger}_3\hat{b}_3+H_{13}+H_{01}$, where \[
H_{13}=\frac{1}{2}\chi'\hat{b}_1^{\dagger}\hat{b}_3\hat{a}+h.c.,
H_{01}=\alpha'\hat{b}_1^{\dagger}\hat{b}_0\hat{b}_0 e^{i\omega_1'
t} +h.c. . \] Here, $H_{13}$ represents the two-photon transitions
between $|1\rangle$ and $|3\rangle$ with $\chi'=-\int d^3x
\phi_1^{*}(x)\chi(x)\phi_3(x)$ while $H_{01}$ represents the
Feshbach resonance coupling of states $|0\rangle$ and $|1\rangle$
with $\alpha'=\alpha \int d^3x \phi_1(x)^{*}\phi^2_0(x)$ and
$\omega_1'=\omega_1+(\epsilon_1-2\epsilon_0)$. The operators
$\hat{b}_i$ are bosonic annihilation operators for atoms or
molecules in state $|i\rangle$ and $\hat{a}$ is the annihilation
operator for photons in the cavity field.

The cavity field is damped at a rate $\kappa$ with the damping
being described by the Born-Markov master equation for a damped
harmonic oscillator. The time evolution of the total density
operator, $w(t)$, is then given by $dw/dt=-i[H,w]+\kappa \left(
2\hat{a}w\hat{a}^{\dagger}-\hat{a}^{\dagger}\hat{a}w-
w\hat{a}^{\dagger}\hat{a} \right)$. In the bad cavity limit where
$\kappa \gg |\chi'| \sqrt{N_0}, |\delta|$, where $N_0$ is the
maximum number of molecules in $|1\rangle$ or $|3\rangle$, the
cavity field can be adiabatically eliminated to give a master
equation for the reduced density operator, $\rho=Tr_{\rm
cavity}[w]$, \cite{search1} \begin{equation}
\dot{\rho}=-i[H_{01}+\delta\hat{b}^{\dagger}_3\hat{b}_3,\rho]+\kappa|\beta|^2\left(
\hat{b}_1\hat{b}^{\dagger}_3\rho\hat{b}_3\hat{b}^{\dagger}_1-\hat{b}_3\hat{b}^{\dagger}_1\hat{b}_1\hat{b}^{\dagger}_3\rho
+ h.c.\right) \label{master} \end{equation} and
$\beta=\chi'/2\kappa \ll 1$. The term proportional to
$\kappa|\beta|^2$ in Eq. (\ref{master}) represents the
amplification of molecules in $|3\rangle$. Only the $|1\rangle
\rightarrow |3\rangle$ transition occurs because the photons
emitted into the cavity are lost before they can be reabsorbed.
The first term in Eq.(\ref{master}), $[H_{01},\rho]$, represents
the oscillatory conversion of pairs of atoms into molecules. For
$\kappa|\beta|^2\gtrsim \alpha'$ the coherent oscillations between
atom pairs and quasi-bound molecules are suppressed and the
conversion of atoms into molecules dominates over the reverse
process.

Equation (\ref{master}) can be used to derive the equations of
motion for the expectation values of the amplitudes, $\langle
\hat{b}_i \rangle=Tr[\hat{b}_i\rho]$, and the populations,
$\langle \hat{n}_i \rangle=Tr[\hat{n}_i\rho]$ where
$\hat{n}_i=\hat{b}^{\dagger}_i\hat{b}_i$,
\begin{eqnarray}
\langle{\dot {\hat n}}_3\rangle&=& 2\kappa|\beta|^2\langle
(1+\hat{n}_3)\hat{n}_1 \rangle
\label{pop3} \nonumber \\
\langle {\dot {\hat b}}_1 \rangle &=& -\kappa|\beta|^2\langle
(1+\hat{n}_3)\hat{b}_1 \rangle
-i\alpha'\langle {\hat b}_0{\hat b}_0 \rangle e^{i\omega_1't} \label{amp1} \nonumber \\
\langle {\dot {\hat n}}_1 \rangle &=& -2\kappa |\beta|^2\langle
(1+\hat{n}_3)\hat{n}_1 \rangle -\alpha'(i\langle {\hat
b}_0\hat{b}_0{\hat b}_1^{\dagger}\rangle e^{i\omega_1't} + c.c. )
\label{pop1} \nonumber
\\
\langle {\dot {{\hat b}_0 {\hat b}_0}}\rangle &=&
-2i\alpha'\langle (1+2{\hat n}_0){\hat b}_1\rangle
e^{-i\omega_1't} \label{pairing} \label{pairing} \nonumber
\end{eqnarray} and $\langle {\dot {\hat b}}_3
\rangle=-i\delta\langle \hat{b}_3\rangle +\kappa|\beta|^2\langle
\hat{n}_1 \hat{b}_3\rangle$. Note that it is the pairing field,
$\langle \hat{b}_0\hat{b}_0\rangle$, rather than the the atomic
field, $\langle \hat{b}_0\rangle$, that drives the formation of
molecules in $|1\rangle$. The atomic population is determined by
conservation of total particle number, $d\left(
\langle\hat{n}_0\rangle+2\langle\hat{n}_1\rangle+2\langle\hat{n}_3\rangle
\right)/dt=0$.

These equations do not form a closed set since the expectation
values of the amplitudes and populations couple to the various
cross correlations between the modes such as $\langle
\hat{n}_1\hat{n}_3\rangle$ or
$\langle\hat{b}_0\hat{b}_0\hat{b}_1^{\dagger}\rangle$. The
simplest approximation is to assume that the correlations
factorize at all times, resulting in the closed set of
differential equations, \begin{eqnarray}
{\dot n}_3 &=& 2(1+n_3)n_1 \label{pop3_factor}\\
{\dot b_1} &=& -[(1+n_3)+\bar{\Gamma}/2]b_1-i\bar{\alpha}P_0e^{i\bar{\omega}_1\tau} \\
{\dot n_1} &=&
-2[(1+n_3)+\bar{\Gamma}/2]n_1-\bar{\alpha}(iP_0b_1^*
e^{i\bar{\omega}_1\tau} +c.c)
\label{pop1_factor} \\
{\dot P}_0 &=& -2i\bar{\alpha}(1+2n_0)b_1 e^{-i\bar{\omega}_1 \tau} \label{pair_factor}\\
{\dot n_0} &=& 2\bar{\alpha}(iP_0b_1^* e^{i\bar{\omega}_1 \tau}
+c.c.) \label{pop0_factor} \end{eqnarray} where $n_i=\langle
\hat{n}_i\rangle$, $b_i=\langle \hat{b}_i \rangle$, and
$P_0=\langle \hat{b}_0\hat{b}_0 \rangle$, and the dot now
indicates a derivative with respect to the dimensionless time
$\tau=\kappa|\beta|^2t$ with
$\bar{\alpha}=\alpha'/\kappa|\beta|^2$ and
$\bar{\omega}_1=\omega_1'/\kappa|\beta|^2$. We have also
introduced a phenomenological decay rate, $\bar{\Gamma}$, for the
intermediate state $|1\rangle$ to describe losses due to inelastic
collisions leading to vibrational decay for $\omega_1<0$ or
disintegration into pairs of atoms for $\omega_1>0$. Equations
(\ref{pop3_factor})-(\ref{pop0_factor}) are independent of the
detuning $\delta$. This follows from the assumption that
$|\delta|\ll \kappa$ used to derive Eq. (\ref{master}), which
implies that the two-photon detuning is much less than the
linewidth of the transition.

The equation of motion for the amplitude of the final state,
$b_3$, can be immediately integrated to give
\begin{equation}
b_3(\tau)=\exp[\int_0^{\tau}d\tau'(-i\delta/\kappa|\beta|^2+n_1(\tau'))]b_3(0).
\end{equation}
Consequently, phase coherent amplification of
$|3\rangle$ only occurs when there is an initial signal to be
amplified, $b_3(0)\neq 0$. Otherwise one only has amplification of
the population as given by Eq. (\ref{pop3_factor}). This remains
true even for the exact dynamics given by (\ref{master}). For the
remainder of this paper we let $\delta=0$.

Consider first the case $\bar{\omega}_1=\bar{\Gamma}=0$ where
$|1\rangle$ is initially in the vacuum state and $n_3\ll 1$. Then
for very short times such that the atomic condensate can be
treated as an undepleted source with $P_0(0)=n_0(0)=2N$ the
population in $|3\rangle$ grows according to
$n_3(\tau)=\exp[2\bar{\alpha}^2(2N)^2\tau^3/3]-1$. This indicates
that even for very short times the amplification cannot be
described via a simple linear rate equation of the form
$dn_3/dt=Gn_3$ where $G={\rm const.}$ For longer times, $b_1$ and
$n_1$ both go to zero while $n_3(\tau)$ saturates at a value that
is less than its maximal value of $n_3(0)+N$, which implies less
than $100\%$ conversion of atoms into molecules (but approaching
$100\%$ for $N\rightarrow \infty$). Fig. 1 shows the dynamics of
the populations and amplitudes for the three modes. The population
and amplitude in $|3\rangle$ grow monotonically while states
$|1\rangle$ and $|0\rangle$ initially oscillate at a frequency
$\sim \bar{\alpha}\sqrt{N}$.

\begin{figure}
\includegraphics*[width=8cm,height=8cm]{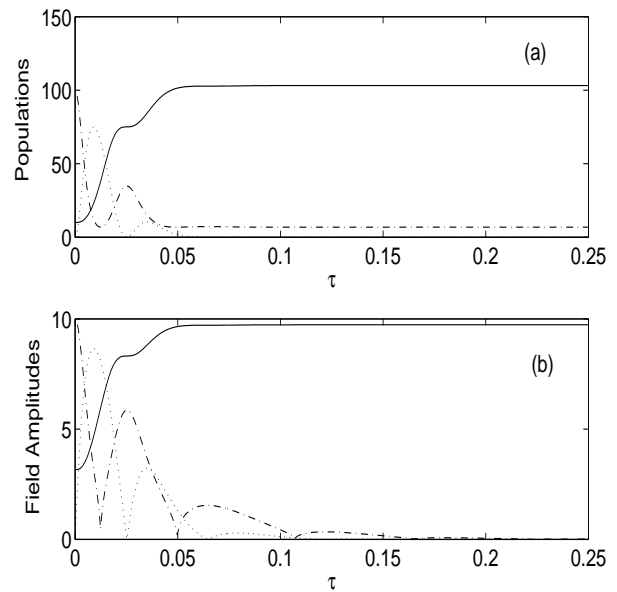}
\caption{(a) Populations of the three modes: $n_3$ (solid line), $n_1$ (dotted line), and $n_0/2$ (dashed dot line).
(b) Amplitudes of the three modes: $|b_3|$ (solid line), $|b_1|$ (dotted line), $\sqrt{|P_0|/2}$ (dashed dot line).
In both plots, $\bar{\alpha}=10$, $\bar{\omega}_1=\bar{\Gamma}=0$, $b_1(0)=n_1(0)=0$, $n_3(0)=b_3(0)^2=10$, and $n_0(0)=P_0(0)=2N=200$.}
\end{figure}

Figure 2 shows the coherent amplification of $b_3$ and the
amplification of the population $n_3$, which includes both
coherent and incoherent contributions, for
$P_0(0)=n_0(0)=2N=500$, $\bar{\alpha}=1$, and various values of
$\bar{\omega}_1$ and $\bar{\Gamma}$. As long as
$\bar{\omega}_1\lesssim \bar{\alpha}\sqrt{N}$, the
$|0\rangle\rightarrow |1\rangle$ transition is resonant and a
finite $\omega_1$ has no significant effect on the dynamics. For
$\bar{\omega}_1\gg \bar{\alpha}\sqrt{N}$ the transition is
off-resonant and this significantly slows the rate at which
population builds up in $|3\rangle$. However, the value at which
the population in $|3\rangle$ saturates is unaffected by
$\bar{\omega}_1$. Decay from the intermediate state only has a
significant effect on the population and amplitude in $|3\rangle$
for $\bar{\Gamma}/2\gtrsim 1+n_3(0)$. In this limit the decay out
of $|1\rangle$ dominates over the initial stimulated scattering
from $|1\rangle$ to $|3\rangle$. The fact that the dynamics are independent
of $\delta$ (for $|\delta|\ll \kappa$) and only weakly dependent on $\bar{\omega}_1$
indicate that mean-field energy shifts due to two-body interactions between atoms
and molecules will have a negligible impact on the amplification of $|3\rangle$.

\begin{figure}
\includegraphics*[width=8cm,height=8cm]{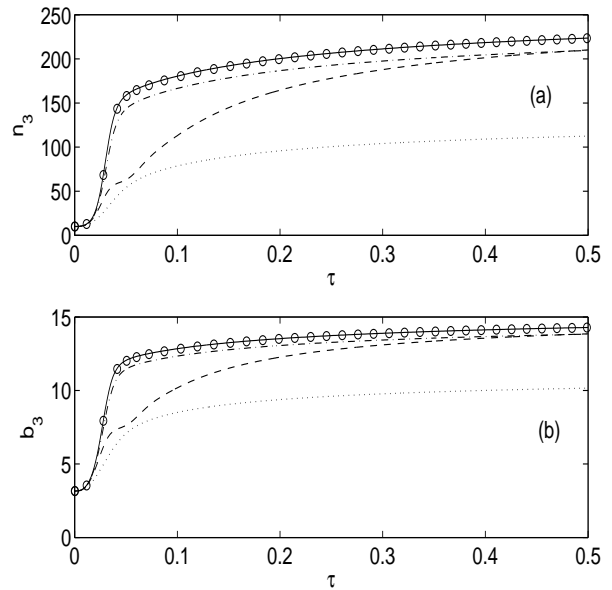}
\caption{(a) Population, $n_3(\tau)$, and (b) amplitude, $b_3(\tau)$,
of the ground state molecular field for the initial conditions
$n_3(0)=b_3(0)^2=10$, $n_0(0)=P_0(0)=500$, and $n_1(0)=b_1(0)=0$.
Solid line: $\bar{\omega}_1/\bar{\alpha}=\bar{\Gamma}=0$; Circles: $\bar{\omega}_1/\bar{\alpha}=10$ and $\bar{\Gamma}=0$;
Dashed line: $\bar{\omega}_1/\bar{\alpha}=100$ and $\bar{\Gamma}=0$; Dashed dot line: $\bar{\omega_1}/\bar{\alpha}=\bar{\Gamma}=10$;
Dotted line: $\bar{\omega}_1/\bar{\alpha}=10$ and $\bar{\Gamma}=100$.}
\end{figure}

We can compare the solution obtained by integrating Eqs.
(\ref{pop3_factor})-(\ref{pop0_factor}) with the expectation
values obtained from direct integration of the master equation,
(\ref{master}), for small numbers of molecules ($\sim 10$). This
is shown in Fig. 3. For Eq. (\ref{master}) we choose the initial
conditions to be a Fock state with occupation number $2N$ for mode
$|0\rangle$, and $|3\rangle$ to be in an arbitrary superposition
of $0$ or $1$ molecules,
$c_0|vacuum\rangle+c_1\hat{b}_3^{\dagger}|vacuum\rangle$. For the
factorized equations motion we then have the corresponding initial
conditions $n_3(0)=|c_1|^2$, $b_3(0)=c_0^*c_1$, and
$n_0(0)=P_0(0)=2N$.

\begin{figure}
\includegraphics*[width=8cm,height=8cm]{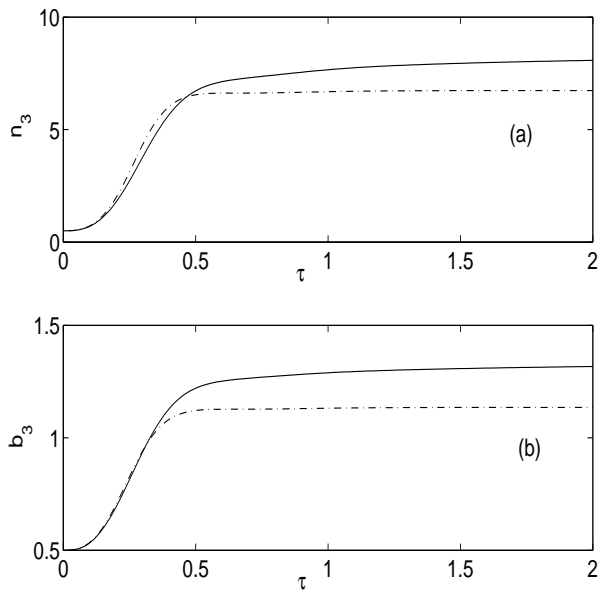}
\caption{(a) Population of the ground state molecular field, $n_3(\tau)$,
calculated using Eq. (2) (dashed dot line) and directly from the master
equation, Eq. (1), (solid line). (b) amplitude of ground state molecules, $b_3(\tau)$,
from Eq. (7) (dashed dot line) and master equation (solid line). In both figures
we have taken the initial amplitudes for state $|3\rangle$ to be $c_0=c_1=1/\sqrt{2}$,
$\bar{\alpha}=1$, $\bar{\omega}_1=\bar{\Gamma}=0$, and $2N=16$ to be the initial population
in $|0\rangle$ with $|1\rangle$ initially in the vacuum state.}
\label{fig3}
\end{figure}

For short times, when the population and field amplitude in
$|3\rangle$ are growing exponentially, the two solutions show good
agreement. For longer times, $\tau\gtrsim 1/2$, the two solutions
for $n_3$ and $b_3$ saturate at different values. Numerical
integration of Eq. (\ref{master}) indicates that the conversion of
atoms into ground state molecules asymptotically approaches
$100\%$ as $\tau\rightarrow \infty$ for all $N$ in contrast to the
results obtained from Eqs.
(\ref{pop3_factor})-(\ref{pop0_factor}). This difference can be
explained by noting that $n_3(\tau)$ and $b_3(\tau)$ saturate when no more new
population forms in $|1\rangle$. From the equation of motion for $\langle \hat{n}_1\rangle$
one can see that the population in $|1\rangle$ is driven by the
cross-correlation $\langle
\hat{b}_1^{\dagger}\hat{b}_0\hat{b}_0\rangle$. When this goes to
zero, no more population is transferred from $|0\rangle$ to
$|1\rangle$ and as a result $n_3(\tau)$ saturates. Figure 4 shows
$\langle \hat{b}_1^{\dagger}\hat{b}_0\hat{b}_0\rangle$ calculated
directly from the master equation and $b_1^*P_0$ calculated using
Eqs. (\ref{pop3_factor})-(\ref{pop0_factor}). As one can see,
$b_1^*P_0$ decays to zero much faster than
$\langle\hat{b}_1^{\dagger}\hat{b}_0\hat{b}_0\rangle$.
Consequently the buildup of population in $|1\rangle$ as given by
Eq. (\ref{pop1_factor}) stops well before that given by the exact
quantum dynamics. We note that other cross-correlations involving
the intermediate state such as $\langle \hat{n}_3\hat{n}_1\rangle$
exhibit similar behavior with the uncorrelated factorized values
obtained from Eqs. (\ref{pop3_factor})-(\ref{pop0_factor})
decaying to zero much faster than the exact values obtained from
the master equation.

In conclusion we have analyzed a model for a coherent matter-wave
amplifier of ground-state molecules in an optical cavity by
analyzing semiclassical rate equations and comparing them to the
exact quantum dynamics.

\begin{figure}
\includegraphics*[width=8cm,height=4cm]{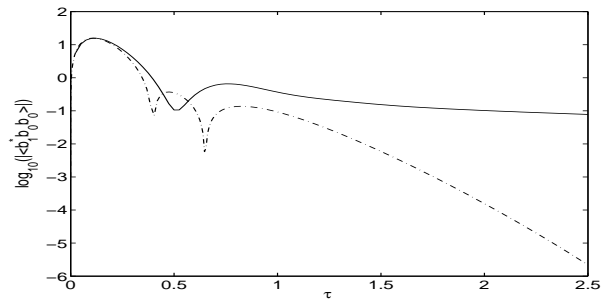}
\caption{$\langle\hat{b}_1^{\dagger}\hat{b}_0\hat{b}_0\rangle$ calculated from the master
equation (solid line) and $b_1^*P_0$ calculated from factorized equations of motion (dashed dot line).
All parameters are the same as Fig. 3.}
\end{figure}

This work is supported in part by the US Office of Naval Research,
the NSF, the US Army Research Office, NASA, and the Joint Services Optics Program.

\end{document}